\documentstyle[11pt,newpasp,twoside,epsf]{article}
\def\edcomment#1{\iffalse\marginpar{\raggedright\sl#1\/}\else\relax\fi}
\def\simgreat{\lower2pt\hbox{$\buildrel {\scriptstyle >}
   \over {\scriptstyle\sim}$}}
\def\simless{\lower2pt\hbox{$\buildrel {\scriptstyle <}
   \over {\scriptstyle\sim}$}}
\reversemarginpar

\begin{document}

\title{{\it Chandra} and {\it RXTE} Observations of X-ray Novae}

\author{Jeffrey E. McClintock}

\affil{Harvard-Smithsonian Center for Astrophysics, 60 Garden Street,
Cambridge, MA 02138, U.S.A.}

\begin{abstract}
We discuss new observations of X-ray novae with {\it Chandra} which
provide strong evidence that black holes have event horizons.  The
evidence is based on the finding that black hole X-ray novae in
quiescence are approximately 100 times fainter than equivalent neutron
star X-ray novae.  The advection-dominated accretion flow model
provides a natural explanation for this difference.  {\it RXTE}
observations of XTE J1550-564 in the {\it very high} state, which were
obtained during the 1998-1999 outburst of the source, reveal an
extraordinarily tight correlation between the central frequency of the
low frequency QPO and the soft, non-power-law flux in the 2-20 keV
band.  We discuss the nature of this soft spectral component and
suggest the importance of obtaining direct observations of it at low
energies ($E~<~2$~keV) at the first available opportunity.
\end{abstract}

\section {Memories of Jan}

I collaborated on several papers with Jan in the 1970s and 1980s.  In
1991 we began to work together on a review chapter for the book
``X-ray Binaries.''  Then suddenly my 17-year-old daughter was killed
in an automobile accident.  I was unable to work for many
months. Finally, I became concerned about the schedule for the review
chapter.  Jan somehow convinced me not to worry at all -- that I
should return to work only when I felt ready.  Meanwhile, he was under
intense pressure, since he was writing one chapter himself,
collaborating on yet another, and helping to edit the entire volume as
well.  Eventually we finished our chapter, and during the whole time
he never pressured me at all.  I am extremely grateful for his
compassion and support.

On another occasion, I experienced a different aspect of Jan's
character.  A collaborator and I had referred to some work of Jan's in
a sloppy way, without giving him and his students proper credit.  Jan
immediately sent us a bristling email that read: ``We already said
that!  We're not just fooling around here!''  In this case, Jan was
being honest, direct and effective.  I learned a lot more than
astrophysics from Jan.

My talk is in three parts.  First, I briefly describe black hole X-ray
novae and neutron star X-ray novae.  Second, I summarize the evidence
that the black hole primaries in X-ray novae possess event horizons.
Finally, I discuss the spectra of black hole X-ray novae at
$E~\simless~2$~keV.

\section {Black Hole X-ray Novae and Neutron Star X-ray Novae}

X-ray novae are characterized by episodic outbursts at X-ray, optical
and radio frequencies, which are separated by long intervals (years to
decades) of quiescence (Tanaka \& Shibazaki 1996; van Paradijs \&
McClintock 1995).  The outburst is caused by a sudden dramatic
increase in the rate of mass accretion onto the compact primary.  The
X-ray flux rises on a time scale of the order of days, and the
subsequent decline of the flux occurs on a time scale of weeks or
months.  The X-ray flux in outburst can be several million times the
quiescent X-ray flux.

For a quiescent X-ray nova, the absorption-line velocities of the
secondary star can be determined precisely.  These velocity data {\it
vs.} orbital phase determine the value of the mass function, $f(M)$,
which gives an absolute lower limit on the mass of the compact
primary.  For 11 systems, $M_{1} > f(M)\ \simgreat\ 3M_{\odot}$.
Assuming that general relativity applies, the maximum mass of a
neutron star can be calculated to be about $2-3M_\odot$ (Cook et
al. 1994; Kalogera \& Baym 1996), so we can be almost certain that the
compact primaries in these 11 systems are black holes.  In addition
the dynamical evidence is strong that two other systems (GRO 0422+32
and 4U1543-47) also contain black hole primaries.  The masses and
other dynamical data for these black-hole X-ray novae are discussed by
Phil Charles in this volume.

A number of X-ray novae also contain neutron star primaries, as
evidenced by the observation of type I X-ray bursts (Lewin, van
Paradijs, \& Taam 1995).  A type I burst, which is a firm signature of
a neutron star, is due to a thermonuclear flash in material that has
been freshly accreted onto the star's surface.  The phenomenology of
these bursts has been closely studied since their discovery in 1975.
The burst rise times are $\sim1-10$~s and their decay times are
$\sim$10~s to minutes.

\section {Evidence for Black Hole Event Horizons from Chandra}

As discussed above, the best evidence for black holes comes from the
dynamical studies of X-ray binaries that contain massive
($M_{1}~\simgreat~3M_{\odot}$) compact stars.  Unfortunately this
evidence is not decisive; in particular, the limiting mass of a
neutron star depends on the presumption that general relativity is the
correct theory of strong gravity.  If we hope to advance our position
further, we must attempt to detect relativistic effects that are
unique to compact objects.  In this context, it would be especially
important to show that one of the dynamical black hole candidates has
an {\it event horizon}, which is the defining property of a black
hole.  Ramesh Narayan, Mike Garcia and I have pursued this goal for
the past several years.  Our approach has been to compare black hole
X-ray novae and neutron star X-ray novae under similar conditions, to
show that their luminosities differ greatly, and to interpret this
luminosity difference in terms of advection-dominated accretion.  In
this section, I sketch this evidence for the detection of an event
horizon.  For a full account of this work, I refer you to a
contemporaneous review by Narayan, Garcia \& McClintock (2001;
hereafter NGM01).

\bigskip

\vbox{
{\small
\begin{center}
{\bf Table~1:~QUIESCENT LUMINOSITIES OF NSXN AND BHXN} \\
\medskip
\begin{tabular}{lcc} \hline \hline
\\
System & $P_{\rm orb}$ (hr) & $\log [L_{\rm min}]~({\rm erg~s^{-1}})$
\\ \\ (1)
& (2) & (3) \\
\\ \hline \\
$\circ$ {\bf SAX J1808.4-3658} & 2.0  & {\bf  31.5} \\
$\circ$ EXO 0748-676    & 3.82           & 34.1    \\
$\circ$ 4U 2129+47      & 5.2              & 32.8    \\
$\circ$ {\bf H1608-52}  &{\bf 12}   & 33.3   \\
$\circ$ Cen X-4         & 15.1      & 32.4   \\
$\circ$ Aql X-1         & 19       & 32.6   \\
   
\\
\hline
\\

$\bullet$ {\bf GRO J0422+32}    & 5.1   &{\bf  30.9},$<31.6$   \\
$\bullet$ {\bf A0620--00}       & 7.8   &{\bf 30.5},30.8   \\
$\bullet$ {\bf GS~2000+25 }     & 8.3   &{\bf 30.4},$<32.2$    \\
$\bullet$ GS1124-683            & 10.4  &$< 32.4$     \\
$\bullet$ H1705-250             & 12.5  &$< 33.0$     \\
$\bullet$ {\bf 4U 1543--47}             & 27.0  &$<${\bf 31.5},$<33.3$
\\
$\bullet$ {\bf GRO J1655--40}           & 62.9  &{\bf 31.3},32.4   \\
$\bullet$ {\bf V404 Cyg}        & 155.3 &{\bf 33.7},33.1    \\

\\
\hline

\end{tabular}
\end{center}
NOTES --- For references see Menou et al. (1999) and NGM01.  New and/or
 {\it Chandra\/} Measurements are {\bf in bold face}.
(1) $\circ$ indicates a neutron star primary and $\bullet$ a black
 hole primary.  (2) Orbital period. (3) Luminosity in
quiescence in the 0.5-10 keV band.\hfill\\
}
}

\subsection {Quiescent Luminosities of X-ray Novae}

In order to fairly compare neutron star X-ray novae (NSXN) and black
hole X-ray novae (BHXN), we must know their orbital periods, as this
has a large influence on their mass transfer rates (Menou et
al. 1999).  In searching for differences between NSXN and BHXN, we
have focused on X-ray observations of these systems in quiescence.
We have examined the quiescent rather than the outburst state because
we expect that the event horizon would be most apparent in quiescence.
We are led to this expectation by the advection-dominated accretion
flow (ADAF) model described below.

While there are now more than a dozen dynamically confirmed BHXN and a
number of NSXN with known orbital periods, sensitive observations of
the quiescent X-ray luminosity exist for only the subset of 8 BHXN and
6 NSXN listed in Table 1.  We note that the tabulated luminosities
assume an X-ray spectral shape and an estimate of the hydrogen column
density to the source; however, the derived luminosities are weakly
dependent on these assumptions (NGM01).

The key observational evidence for event horizons is summarized
in Figure~1, where the Eddington-scaled quiescent luminosities of
NSXN and BHXN are shown as a function of the binary orbital period
$P_{orb}$.  We scale the luminosity by the Eddington luminosity
$L_{\rm Edd}$ under the reasonable assumption that the
Eddington-scaled mass accretion rate is similar in quiescent BHXNs and
NSXNs (NGM01).  For the orbital period range in question, this
assumption has been confirmed by model calculations (Menou et
al. 1999).

\begin{figure}
\plotfiddle{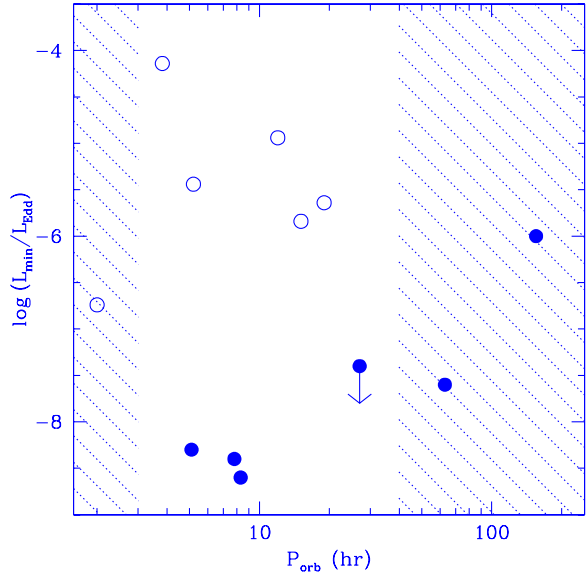}{4.0in}{0.}{40.}{40.}{-112}{+13}
\vspace{-0.9in}
\caption{Quiescent luminosities of BHXN (filled circles) and NSXN
(open circles) plotted against the binary orbital period.  Only the
lowest quiescent detections, or Chandra upper limits, are shown.  The
non-hatched region includes both BHXN and NSXN and allows both kinds
of X-ray novae to be compared directly.  The hatched region on the
left has no BHXN and the region on the right has no NSXN; therefore
these regions of the plot are less useful.}
\end{figure}

Versions of Figure~1 were presented in earlier papers (e.g. Narayan,
Garcia \& McClintock 1997; Menou et al. 1999), and
in each case we claimed that the data indicated that BHXN are
substantially dimmer than NSXN in quiescence.  The version shown here,
which is taken from Garcia et al. (2001) and includes a number of
sensitive BHXN measurements made with {\it Chandra}, greatly
strengthens our earlier claims.  We see that BHXN are dimmer than NSXN
with comparable orbital periods by a factor of 100 or more (see the
non-hatched region of the plot).  Such a gross difference implies an
important qualitative difference in the nature of the accretors in the
two kinds of systems--e.g. the presence of an event horizon
{\it~versus} a surface, as discussed below.

At the shorter orbital periods, ($P_{orb}$~\simless~1~day), the rate
of mass transfer is rather low and is driven largely by gravitational
radiation.  Two BHXN are located at long orbital period in the
cross-hatched region of Figure 1.  For these systems, V404 Cyg and GRO
1655-40, nuclear evolution rather than gravitational radiation is
expected to drive the mass transfer at a high rate, which is no longer
uniquely determined by $P_{orb}$.  Furthermore, there are no NSXN with
comparably long values of $P_{orb}$. For both reasons these particular
BHXN are less useful for comparison.  One NSXN with a very short
orbital period, SAX J1808.4-3658, lies in the cross-hatched region on
the left.  It is also not useful since there is no comparison BHXN
with a comparable period.  This NSXN is also anomalous in being the
only known millisecond X-ray pulsar (Wijnands et al. 1998).

\subsection {Advection-Dominated Accretion and X-ray Binaries}

Apart from the thin accretion disk, a second stable accretion flow
solution is known, the so-called advection-dominated accretion flow
(ADAF).  The key feature of an ADAF (see NGM01, and references
therein) is that the heat energy released by viscous dissipation is
not radiated immediately, as in a thin disk, but is stored in the gas
as thermal energy and advected with the flow--hence the name ADAF.
Thus, as the gas falls into the deep potential well of the compact
object, it becomes extraordinarily hot, $T_{i}~\sim~10^{12}$K/r, where
r is the radius in Schwarzschild units.  At such temperatures, the gas
bloats up around the compact object in a quasi-spherical cloud of low
density gas.

The low density of an ADAF is a key property, since it reduces the
number of particle-particle interactions, which thereby makes the
plasma radiatively inefficient.  The low density is also largely
responsible for creating a two-temperature plasma, with the electrons
being cooler than the ions, though still quite hot:
$T_{e}~>~10^{9}$~K.  Heat energy from viscous dissipation is assumed
to go primarily to the ions, which are inefficient radiators.  Energy
must be transferred from the ions to the electrons before it can be
radiated, but that process is slow and the gas reaches the compact
object first.  Both because the energy is bottled up in the ions and
because the radiative efficiency of the electrons is poor
at low densities, the gas becomes advection-dominated.

Detailed calculations (Narayan \& Yi 1995) show that the
optically-thin two-temperature ADAF solution is allowed only for
Eddington-scaled mass accretion rates $\dot m\equiv\dot M/\dot M_{\rm
Edd}$ less than a critical value $\dot m_{\rm crit}\sim 0.01-0.1$.
The X-ray spectra computed with the ADAF model match observations of
BHXN in the {\it low hard} state quite well (Esin et al. 1997, 1998).
Moreover, the prediction of a central hole in the thin disk that is
filled with hot ADAF gas (Narayan, McClintock \& Yi 1996; Narayan
1996; Esin et al. 1997) has been confirmed by recent observations of
the BHXN XTE~J1118+480 (McClintock et al. 2001; Esin et al. 2001;
\S4.2): the hole in the disk must have a radius
$\simgreat~55R_{\rm~Schw}$, rather than the $3R_{\rm~Schw}$ observed
in the bright outburst state (see \S4).

According to the above scenario, as the mass accretion rate decreases,
the thin disk recedes to larger and larger radii and its luminosity
falls rapidly.  The emission from the ADAF also decreases
substantially because the radiative efficiency decreases with
decreasing $\dot m$ (due to the lower density).  The luminosity of the
source thus falls steeply; roughly $L\sim \dot m^2$ (Narayan \& Yi
1995).  At sufficiently low $\dot m$, the luminosity corresponds to
that observed in the quiescent state of X-ray novae, and in this state
the ADAF zone is expected to be quite large ($\sim10^3-10^4R_{\rm~Schw}$
in models).  

The ADAF model was first applied to the quiescent state of BHXN.
Narayan, McClintock \& Yi (1996) argued that the BHXN A0620-00 in
quiescence has a spectrum that is inconsistent with a thin disk model,
and showed that the spectrum can be explained with a model that has an
ADAF at small radii and a thin disk at large radii.  Narayan, Barret
\& McClintock (1997) modeled the relatively high signal-to-noise
spectrum of the BHXN V404 Cyg and showed that the model again fits the
observations quite well.  Hameury et al. (1997) showed that the 6-day
X-ray delay relative to the optical observed by Orosz et al. (1997)
during the onset of an outburst of the BHXN GRO 1655-40 is best
understood by invoking a hole in the thin disk, as postulated in the
ADAF model.  

\subsection {The Event Horizon}

The ADAF model plus the event horizon provide a straightforward
explanation for the relative faintness of BHXN in quiescence.  The
reasoning is simple and based on the key feature of an ADAF: namely
that most of the energy remains locked in the gas and is advected to
the center.  For quiescent X-ray novae, only a tiny fraction is
radiated (less than $10^{-2}$ and as low as $10^{-4}$ in some models,
compared to $\sim~10^{-1}$ for thin disk accretion).  What happens to
the bulk of the thermal energy, $\sim0.1\dot Mc^2$, when it arrives
at the compact object?

If the accretor is a black hole, the advected energy disappears
through the horizon as the gas falls in.  But if the accretor is a
neutron star, the gas will heat the surface of the star and radiate
all its stored energy.  Thus, for accretion via an ADAF the luminosity
of a neutron star, or any object with a surface, will be significantly
larger than an object with an event horizon (Narayan \& Yi 1995;
Narayan, Garcia \& McClintock 1997; Garcia et al. 2001; NGM01).  The
data (Fig. 1) show clearly that quiescent BHXN are much dimmer than
quiescent NSXN.  Since we expect the two kinds of sources to have
similar Eddington-scaled accretion rates (see above), we interpret the
observations as the first clear evidence for the presence of event
horizons in black hole candidates.

\subsection {Further Developments in Theory and Modeling}

The ADAF model discussed above is qualitatively consistent with the
data in Figure 1; however, the situation is not as satisfactory on a
quantitative level.  The above model predicts that the difference in
luminosity between BHXN and NSXN should be much larger than a factor
of $\sim100$ seen in the data.  Menou et al. (1999) attribute this
discrepancy to the NSXN, which they argue are anomalously dim because
of the propeller action of a spinning magnetized neutron star (see
NGM01).

ADAFs are convectively unstable (Narayan \& Yi 1994, 1995).  ADAFs
with convection, which are called convection-dominated accretion
flows, or CDAFs, are a new and important area of study (NGM01).
Whether accretion proceeds via an ADAF or a CDAF, the mass that does
reach the compact object will arrive at nearly the virial temperature
with a great deal of thermal energy.  Consequently, a BHXN will be
significantly dimmer than an NSXN.  Thus the interpretation of Fig. 1
as evidence for the event horizon is valid for either an ADAF or a
CDAF.  However, since the radiative efficiency of a CDAF is higher
(for the same mass accretion rate), the difference in luminosity
between BHXN and NSXN will be smaller than that estimated by Menou et
al. (1999), who assumed an ADAF.  In fact, Abramowicz \& Igumenshchev
(2001) have argued that the factor of 100 difference shown in Figure 1
between BHXN and NSXN is consistent with the CDAF model.  Thus, the
CDAF model provides a better quantitative match between model
predictions and observations of the luminosity, thereby strengthening
the case for the detection of the event horizon.  An important next
step will be the computation of model spectra to compare with
observations (e.g. Kong et al. 2001).

We have interpreted the data summarized in Figure~1 in terms of the
ADAF model--i.e. we have assumed that X-ray novae are powered by
accretion.  However, several alternative models have been proposed to
explain the difference in X-ray luminosity between BHXN and NSXN, most
notably those by Bildsten and Rutledge (2000) and Brown, Bildsten \&
Rutledge (1998).  For a critical discussion of these and other
alternative models see NGM01.

\section{X-ray Spectra of BHXN at Energies Below $\sim$2 keV}

The quest to detect a black hole event horizon may seem somewhat
inflated as a topic, so I have picked something quite modest to end
with.  Namely, the spectra of accreting black holes at energies below
$\sim$2~keV.  The flavor of this topic is similar to many other topics
that Jan and I discussed over the years.

\subsection{The High/Soft State}

For BHXN in the {\it high/soft} state, the spectrum can be very well
modeled by a multi-temperature disk blackbody (Mitsuda et al. 1984;
Makashima et al. 1986).  In this state, there is good evidence that
the accretion disk extends inward to the minimum stable circular orbit
at $R_{\rm ms}~=~3R_{\rm~Schw}$ (Tanaka \& Shibazaki 1996; Sobczak et
al. 1999).  The low energy spectrum of a BHXN in the {\it high/soft}
state may approximate the model disk spectrum fairly well; however,
because of interstellar absorption no sensitive observations have been
made at low energies ($E~<$~1~keV) that confirm this conjecture.

\subsection{The Low/Hard State of XTE J1118+480}

For a BHXN in the {\it low/hard} state, it has been believed for some
time that the inner edge of the accretion disk is located far outside
the minimum stable orbit, which is located at $3R_{\rm~Schw}$.  The
presence of a large hole in the inner disk was predicted by ADAF
theory (Narayan, McClintock \& Yi 1996; Narayan 1996; Esin, McClintock
\& Narayan 1997), and further support for this picture was provided
subsequently by Compton reflection models that were constructed for
several sources (e.g. Gierlinski et al. 1997; Zycki, Done \& Smith
1998).

In March of last year, XTE J1118+480 erupted.  The source's
high-latitude ($b~=~62^{\rm o}$) and extraordinarily low column depth,
$N_{\rm H}~\approx~1 \times 10^{21}$ cm$^{-2}$, gave us our first
penetrating look at the low-energy spectrum of a BHXN in the {\it
low/hard} state.  Two major multiwavelength studies were conducted, one by
Hynes et al. (2000) and the other by McClintock et al. (2001).  Both
studies feature near-simultaneous radio, UKIRT IR, HST/STIS, EUVE and
RXTE data.  In addition, the latter study also includes simultaneous
Chandra LETG/ACIS grating data that covers the energy range
0.24--6~keV.

For a global view of the spectrum, which consists of two components,
see Figure~4 in McClintock et al. (2001).  The quasi power-law
spectral component, which was observed from 0.4--160~keV, was modeled
successfully as an ADAF (Esin et al. 2001).  Below $\sim0.4$ keV, the
spectrum is dominated by a $\approx$24~eV thermal component that is
attributed to an accretion disk with a large inner disk radius:
$\simgreat~55R_{\rm~Schw}$ (Esin et al. 2001), which is about 20 times
larger than the radius of the minimum stable orbit (3$R_{\rm~Schw}$).
In conclusion, the low energy spectrum of the {\it low/hard} state has
been closely observed.

\subsection{The Very High  State of XTE J1550-564}

Unfortunately, we have yet to observe a high latitude BHXN in the {\it
very high} state.  Nevertheless, I will argue that the extensive
spectral and timing results obtained for XTE~J1550-564 give us some
insight into the {\it very high} state spectrum of this source below
$\sim2$~keV, despite its large column depth, ($N_{\rm~H}~\sim~10^{22}$
cm$^{-2}$ (Sobczak et al. 2000; Tomsick et al. 2001).  A total of 209
pointed observations of XTE J1550-564 were made during its 250-day
outburst in 1998-99.  The spectra are generally well-fitted by a model
comprised of disk blackbody plus power-law components as discussed in
detail by Sobczak et al. (2000).  Low-frequency QPOs were detected in
72 of the 209 observations and fruitfully divided into three types
based on their phase lags and other properties by Remillard et
al. (2001).  Here we consider only the subset of 46 observations that
yielded ``type~C'' low-frequency QPOs.

In the context of the disk blackbody model, Remillard et al. (2001)
examined the correlation between the central frequency of the type~C
QPOs and the flux from the accretion disk.  Because the temperature of
the inner disk of XTE J1550-564 is cool,
(k$T_{\rm~in}~\approx$~0.6~keV), we faced a serious difficulty in
defining the accretion disk flux, since a large fraction of this soft
flux is cut off by the ISM and the PCA response and could not be
observed (see Fig. 2).  We therefore considered two measures of the
disk flux: (1) The observed 2-20~keV flux and (2) the bolometric flux,
$F_{\rm bol}~\propto~R_{\rm in}^{2}~\times~T_{\rm in}^{4})$, where
$R_{\rm in}$ is the inner disk radius (Mitsuda et al. 1984; Makashima
et al. 1986).  It was not obvious to us which of these measures of the
disk flux one should correlate against the QPO frequency, so we
prepared the two correlation plots shown in Figure 3 (Remillard et
al. 2001; see their Fig.~5 for additional correlations between QPO
parameters and spectral parameters).

\begin{figure}
\plotfiddle{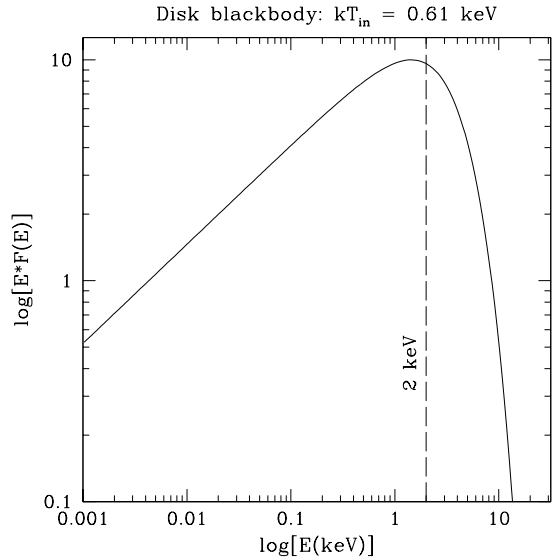}{4.0in}{0.}{38.}{38.}{-112}{+30}
\vspace{-0.9in}
\caption{A disk blackbody model spectrum generated using XSPEC
(diskbb) for the average disk temperature determined by Sobczak et
al. (2000) for the 46 observations that exhibited type C QPOs
(Remillard et al. 2001).  Note that the quantity on the y-axis is
$(Energy~\times~Energy~flux)$.  The dashed vertical line indicates the
approximate low-energy cutoff of the RXTE PCA and the ISM}
\end{figure}

\begin{figure}
\plotfiddle{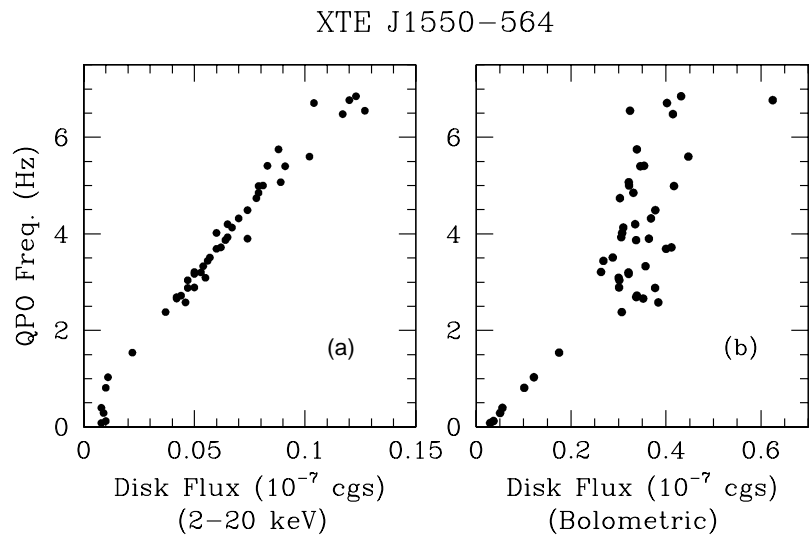}{6.0in}{0.}{65.}{65.}{-202}{+60}
\vspace{-3.2in}
\caption{QPO central frequency for Type C QPOs {\it vs.} the disk flux
in units of 10$^{-7}$ ergs~cm$^{-2}$~s$^{-1}$. (a) The unabsorbed
flux measured over the observed 2--20 keV band. (b) The bolometric
disk flux.}
\end{figure}

The excellent correlation shown in Figure 3a between an oscillation
frequency and the observed 2-20 keV accretion disk flux strikes me as
extraordinary.  Especially when one considers the enormous amplitudes
of these oscillations, which range from 5-15\% (rms; see Fig. 5 in
Remillard et al. 2001).  How are such powerful oscillations coupled so
effectively to the disk flux?

The correlation of frequency with bolometric flux in Figure 3b is
shabby at the higher frequencies.  This result may be caused by
sizable uncertainties in both $T_{\rm in}$ and $R_{\rm in}$ combined
with the large extrapolation required to compute the bolometric flux.
On the other hand, the model's failure may be caused by the failure of
the disk blackbody model itself, which brings me to my last point.  In
Figure~4, I show a typical {\it very high} state spectrum of XTE
J1550-564.  As shown, in this state the accretion disk component is a
tiny fraction of the total flux, which makes the tight correlation
shown in Figure 3a all the more remarkable.  All along, I have been
referring to this soft component as thermal emission from an accretion
disk (cf. Sobczak et al. 2000).  However, at this symposium Fred Lamb
pointed out to me that this soft flux is more likely a Comptonized
component of emission, which one might expect to find in the presence
of the dominant power-law component (Fig. 4).

\begin{figure}
\plotfiddle{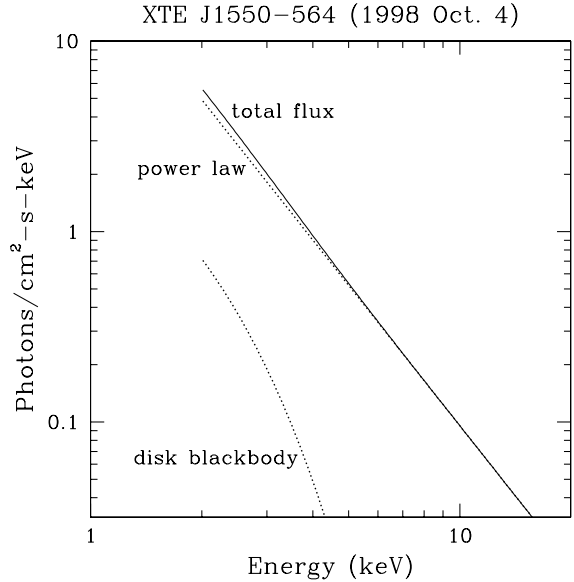}{4.0in}{0.}{39.}{39.}{-112}{+22}
\vspace{-1.1in}
\caption{The solid line shows a typical PCA {\it very high} state
spectrum of XTE J1550-564 obtained when type C QPOs were present
(observation no. 36; Sobczak et al. 2000).  The dashed lines show the
disk blackbody (k$T_{\rm in}$ = 0.64 keV) and power-law (photon index
= 2.45) components of emission.  The statistical uncertainties in the
spectral parameters are very small (see Table 2 in Sobczak et
al. 2000).}
\end{figure}

In conclusion, in an effort to be as objective as possible, I now
refer to this soft spectral component in the {\it very high} state as
the ``non-power-law'' component--i.e. the residual flux that remains
after fitting a pure power-law to the data.  The relative contribution
of this non-power-law component increases with decreasing energy,
reaching $\approx15$\% at 2~keV (Fig. 4).  Clearly it will be
important to capitalize on any rare future opportunities that may
arise to observe the low energy spectrum and other properties of this
soft emission component, which is coupled so strongly to the powerful
low-frequency QPOs.

\smallskip

\noindent {\bf Acknowledgment:} \S3 on black-hole event horizons
draws heavily on a major review by Narayan, Garcia and McClintock
(2001; NGM01).  I am grateful to Ramesh Narayan and Mike Garcia for
permission to make use of this material.  This work was partially 
supported by NASA grants NAG5-10813 and GO0-1105A.

\end{document}